# Topological Defects in Semiconducting Carbon Nanotubes as Triplet Exciton Traps and Single-Photon Emitters

Timur Biktagirov[1*], Uwe Gerstmann[1], Wolf Gero Schmidt[1]

[1]Universität Paderborn, Department Physik, Warburger Str. 100, 33098 Paderborn, Germany

* Timur.Biktagirov@upb.de

**ABSTRACT:** We investigate how topological defects influence the excitonic behavior in (6,5) semiconducting single-walled carbon nanotubes. Our theoretical study demonstrates that topological defects, particularly the widely occurring Stone-Wales defect, can act as efficient traps for triplet excitons, characterized by significant zero-field splitting consistent with experimental data and a small singlet-triplet energy gap. Additionally, the weak electron-phonon coupling positions these defects as promising candidates for single-photon emission at telecom wavelengths (1.6 μm). These findings pave the way for enhancing the performance of carbon nanotube-based quantum light sources and optoelectronic devices.

1. INTRODUCTION

Semiconducting single-walled carbon nanotubes (SWCNTs) are quasi-one-dimensional materials with unique photonic functionalities [1] that can be exploited in optoelectronic devices [2–4], quantum-enhanced sensors [5, 6], and telecom-range single-photon emitters [7–9]. SWCNTs can now be produced with a specific chirality, a structural parameter defined by the (*n*, *m*) indices that describe the rolling vector of the graphene sheet, which governs their electronic structures and optical responses [10, 11]. However, the details of their excited state dynamics, which are imperative for effective photophysical and quantum technological applications, remain a focus of active research. One particularly intriguing aspect is the role of structural defects in their exciton behavior. It is known that excitons (electron-hole pairs) in SWCNTs exhibit large binding energies [12] and very high mobility [13]. Therefore, if the nanotube walls contain defects, fast-diffusing free excitons are likely to be trapped at the defect-induced electronic states and localize at the defect site (see Figure 1A) [14]. Intentional implantation of defects by doping or covalent modifications of semiconducting SWCNTs has been long recognized as a promising approach to increase photoluminescence quantum yields by preventing free excitons from encountering recombination sites, as well as to enable single-photon emission [15–19].

Initially, a defect-trapped exciton (DTE) is formed in a singlet state ($S$ = 0), but it can transition to a triplet state ($S$ = 1) through a spin-forbidden process known as intersystem crossing (ISC) [20]. Photogenerated triplet states are of particular interest because they can be directly detected by electron paramagnetic resonance (EPR) and related techniques, offering unique insights into the material's photophysics. Furthermore, typically long lifetimes and quantum coherence times of triplet excitons in organic materials make them pivotal to light-emitting devices [21] and promising spin qubit candidates [22]. In semiconducting SWCNTs, triplet excitons have been observed using transient absorption spectroscopy [23, 24] and optically detected magnetic resonance (ODMR) [25–28]. Notably, ODMR measurements in (6,5) chirality SWCNTs were able to resolve the definitive spectroscopic fingerprint of triplet spin states – the zero-field splitting (ZFS) between the $m_S = 0$ and $m_s = \pm 1$ spin sublevels [26, 28]. In light-element materials, ZFS arises from the magnetic dipolar interaction between the two unpaired electrons of the triplet and reflects their mutual spatial distribution [29]. The substantial magnitude of the ZFS observed in (6,5) SWCNTs suggests that the triplet excitons are well-localized. Additionally, the work of Sudakov *et al.* [28] identified coexistence of at least two types of photoinduced triplets with distinct ZFS values (550 MHz and 150 MHz) and hints at a role of defects in forming these triplet states, even in nominally pristine nanotubes.

Following these experimental results, we aimed to explore the role of native defects in the formation and localization of excitons in semiconducting SWCNTs. Specifically, we focus on one of the most common classes of intrinsic defects: topological defects (TDs), which are local formations of nonhexagonal carbon rings. TDs can occur intrinsically during the synthesis, purification, or chemical processing of SWCNTs or introduced intentionally [30–32]. A prominent example of TDs is the Stone-Wales (SW) defect, created by a 90-degree rotation of a C-C bond, resulting in two heptagons and two pentagons (5-7-7-5) [33]. SW defects, like many other TDs, do not alter the chirality of nanotubes and are therefore difficult to detect experimentally, often making the nanotubes appear nominally pristine.

In this work, we employ electronic structure calculations based on density functional theory (DFT) and choose the (6,5) SWCNT as a model system because this chirality was used in previous ODMR studies (Refs. [25, 26, 28]). We then apply our recently developed method for calculating magnetic dipolar interactions in extended periodic systems, incorporating spin decontamination corrections [34, 35], to compute the spin-spin ZFS tensors of the corresponding DTEs and compare them with experimental ODMR data. Finally, we investigate electron-phonon interactions in the most notable TD, discussing its potential as a single-photon emitter.

## 2. METHODS

The DFT calculations were carried out in the Quantum ESPRESSO program suite [36, 37] using a supercell model with periodic boundary conditions. The supercell contained 364 carbon atoms, corresponding to approximately a 4.07 nm segment of the nanotube. In both lateral dimensions, the nanotube in the supercell was surrounded by 16 Å of vacuum. Structural relaxation was performed at the gamma point with the plane-wave energy cutoff of 40 Ry and the PBE exchange-correlation functional [38], yielding the C-C bond lengths of 1.429 Å (along the axis) and 1.433 Å (perpendicular to the axis) and the nanotube diameter of about 0.76 nm. This agrees well with the expected C-C bond length, $a_{CC} = 1.44$ Å, and the corresponding diameter conventionally estimated as $a_{CC}\{3(n^2 + nm + m^2)\}/\pi$ = 7.57 Å [1]. Subsequently, the HSE06 [39] hybrid functional was utilized for electronic structure and ZFS calculations. To obtain the ZFS tensor, we first computed the sum of pairwise spin-dipolar interactions between all the occupied Kohn-Sham orbitals $p$ and $q$ in the form of 3 × 3 matrices **d** [34]:

$$d_{ab} = \sum_{p,q} \chi_{pq} \int \frac{|\mathbf{r}-\mathbf{r}'|^2 \delta_{ab} - 3(\mathbf{r}-\mathbf{r}')_a (\mathbf{r}-\mathbf{r}')_b}{|\mathbf{r}-\mathbf{r}'|^5} \times \left[ n_{pp}(\mathbf{r}) n_{qq}^*(\mathbf{r}') - n_{pq}(\mathbf{r}) n_{pq}^*(\mathbf{r}') \right] d\mathbf{r} d\mathbf{r}',$$

where $a$ and $b$ refer to the Cartesian coordinates $x$, $y$, and $z$. The two-electron densities $n_{pq}(\mathbf{r})$ reflect the spatial distribution of the Kohn-Sham orbitals, and the coefficient $\chi_{pq}$ originates from the matrix elements of spin-operators. Since the standard DFT-based calculation of spin-dipolar ZFS is known to suffer from spin contamination, we follow our recently developed approach to minimize its effect [35]. Namely, we define the ZFS tensor as $\mathbf{D} = \frac{\alpha^2}{8}(\mathbf{d}-\mathbf{d}_{BS})$, where $\alpha$ is the fine structure constant, and $\mathbf{d}_{BS}$ is the same as $\mathbf{d}$ but computed for the broken symmetry singlet ($S$ = 0) state of the system. The latter is constructed by constraining the occupations so that one of the unpaired electrons is transferred to its corresponding spin-down orbital.

The excited state calculations were carried out in the framework of constrained occupation DFT (commonly referred to as the ΔSCF approach) [40, 41]. While ΔSCF is widely used to calculate accurate excitation and emission energies of solid-state defects, it describes excited states as single Slater determinants, which may yield inaccurate results if the modeled state has significant multi-determinant character. To verify the validity of the single-determinant approximation, we compared ΔSCF results with time-dependent DFT (TDDFT), which describes excited states as linear combinations of Slater determinants. For this comparison, we isolated a finite-size (~4.3 nm, 388 carbon atoms) SWCNT segment with the defect located at the center, passivated the dangling bonds with hydrogen atoms, and performed TDDFT calculations using an atom-centered basis set (def2-SVP [42]) in the ORCA program (version 6.0) [43]. The segment size was sufficiently large relative to the defect localization to minimize finite-size effects and adequately describe the character of the Kohn-Sham states involved in the studied excitations. The TPSSh meta-hybrid functional [44] was selected for TDDFT due to its good agreement with the HSE06 ground-state electronic structure obtained for the supercell.

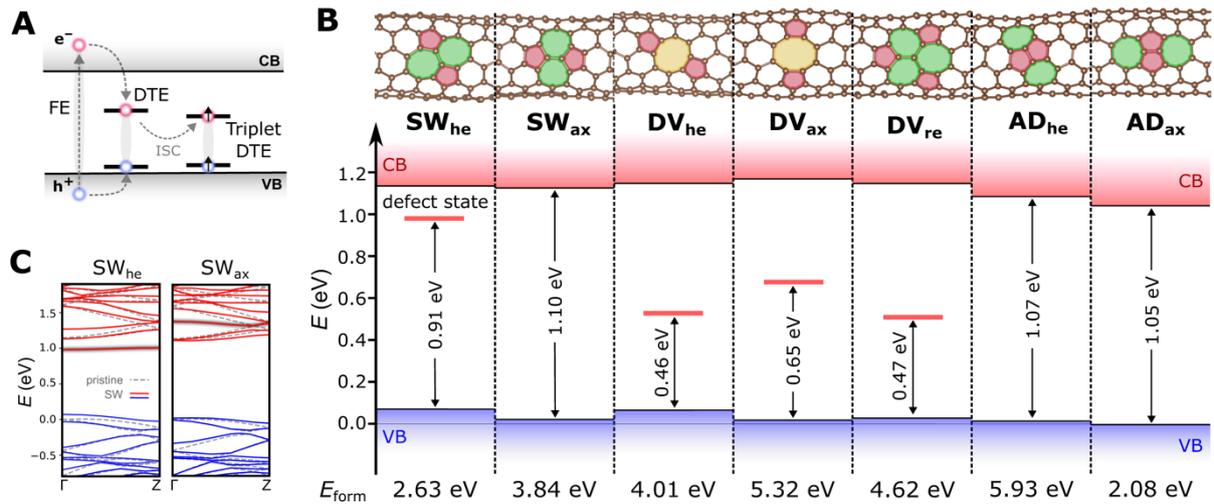

**Figure 1.** (A) Schematic depiction of the formation of a free exciton (FE), a defect-trapped exciton (DTE), and a triplet DTE. (B) Sketch of the atomic structures of the modelled topological defects in the (6,5) SWCNT (*top*) and the corresponding Kohn-Sham states at the Γ-point (blue = occupied states, red = unoccupied states). Defect-induced *unoccupied* states in the band gap are depicted as solid red lines. In case of SW$_{ax}$, AD$_{he}$, and AD$_{ax}$, these states are situated in the conduction band (CB). For all cases, the energy gap between the highest occupied and lowest unoccupied states are marked with vertical arrows. The zero energy is set at the VB maximum of pristine (6,5) SWCNT. (c) Calculated band structures of the SWCNT containing $SW_{he}$ and $SW_{ax}$ (solid lines; blue = occupied, red = unoccupied) superimposed with that of the pristine (6,5) nanotube (dashed lines). Localized defect states are highlighted with dark grey shading.

## 3. RESULTS AND DISCUSSION

We begin our analysis by modeling the (6,5) SWCNT containing the TDs in their singlet ground state. The studied TDs are depicted in Figure 1B. All these defects can be introduced into the SWCNT at the cost of local distortion without the global chirality change (unlike, for example, a simple 5-7 pair [45]). Aside from two SWs with different orientations relative to the SWCNT axis (*helical*, $SW_{he}$, and *axial*, $SW_{ax}$), we considered another common TD — divacancies ($DV_{he}$ and $DV_{ax}$) formed by the removal of two adjacent carbon atoms, resulting in the structure comprised of one octagon and two pentagons (5-8-5) [46]. Since this DV configuration is fully reconstructed to form an sp²-hybridized system, it does not contain dangling bonds (in contrast to a single vacancy). Additionally, the divacancy defect in graphene is reported to spontaneously transform into a stable defect structure consisting of three pentagon and three heptagon rings [47]. We consider this reconstructed divacancy and label it $DV_{re}$.

We also modeled another noteworthy type of TD formed by the adsorption of a carbon dimer [48, 49]. Unlike single adatoms, a dimer can incorporate as a part of the sp²-hybridized carbon system and therefore can be regarded as a TD. The resulting ad-dimer TD (denoted as $AD_{he}$ and $AD_{ax}$ in Figure 1B) comprises two pentagons and two heptagons (7-5-5-7) and is often referred to as an inverse Stone-Wales defect [50].

First, we calculated the formation energies of the TDs as $E_{form} = E_{defect} - E_p + n\mu$, where $E_{defect}$ and $E_p$ are total energy of defect-containing and pristine supercells, respectively, $\mu$ is the chemical potential of carbon estimated as the total energy per atom in pristine nanotube, and $n$ is number of removed/added carbon atoms ($n$ = 0 for SW, 2 for DV, and –2 for AD). The calculated formation energies are listed in Figure 1B and range from 2 to 6 eV, in qualitative agreement with values reported for similar defects in graphene [51, 52]. Our results suggest $SW_{he}$ and $AD_{ax}$ as the most energetically favorable TDs among those considered. There is also a notable difference in the $E_{form}$ values between axial and helical configurations of each TD, which is expected due to induced local deformations and differences in strain distribution patterns.

Next, we analyzed how each TD affects the electronic structure of the SWCNT. The Kohn-Sham states corresponding to the valence band (VB) maximum and conduction band (CB) minimum for defect-containing supercells are depicted in Figure 1B. The incorporation of TDs perturbs the VB and CB states, leading to a slight decrease in the band gap compared to the pristine (6,5) SWCNT, which is 1.15 eV in our DFT calculations. Notably, the $SW_{he}$ defect introduces an additional unoccupied state into the band gap, situated close to the CB minimum (0.15 eV below). When an electron is excited from the VB to the CB (e.g., through photoexcitation), it is likely to fall into this shallow trap (cf. Figure 1A). Similarly, each divacancy features one clearly distinguishable empty defect level in the band gap. However, unlike the shallow trap introduced by $SW_{he}$, divacancy defect states are classified as deep (mid-gap) traps, lying far from the band edges. These deep traps typically act as efficient recombination centers. An electron trapped in a mid-gap state can recombine with a hole from the VB, and vice versa [53, 54].

We then examined the band structure of the two SW defect configurations in more detail (Figure 1C). Despite representing the same defect type, they have strikingly different manifestation in the band structure of the SWCNT. In contrast to SW$_{he}$, the supercell containing SW$_{ax}$ does not exhibit unoccupied states in the band gap. However, the inspection of CB reveals a localized defect-induced state with nearly flat dispersion about 0.3 eV from the CB minimum, which may stabilize upon trapping a photoexcited electron.

Another prominent difference between SW$_{he}$ and SW$_{ax}$ is their effect on the VB maximum. In the pristine (6,5) SWCNT, the VB maximum at the Γ-point comprises two degenerate states (see dashed lines in Figure 1C). These states have similar electron density distributions, delocalized over the entire nanotube [14]. Introducing TDs lifts this degeneracy. In the case of SW$_{ax}$, the resulting splitting at the Γ-point is relatively subtle, below 0.03 eV. In contrast, the effect is substantial in the supercell containing the SW$_{he}$ defect. While one of the former VB maximum states largely retains its character and delocalization, the other shifts up into the band gap by about 0.1 eV and becomes significantly localized around the TD.

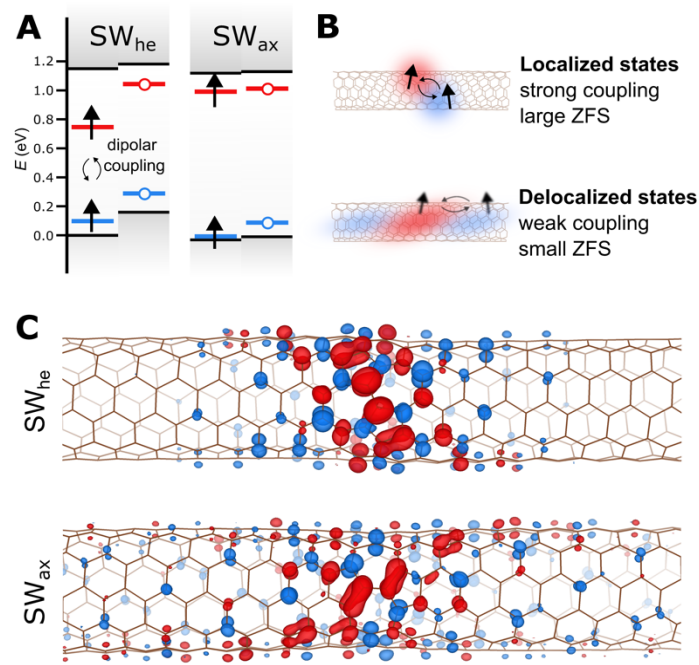

**Figure 2**. (A) Schematic diagram of the DFT calculated electronic structure of the triplet ($S = 1$; DTE) states of SW$_{he}$ (left) and SW$_{ax}$ (right). Arrows represent occupied spin-up states, while empty circles indicate unoccupied spin-down states, collectively constituting the triplet. (B) A sketch illustrating the relation between delocalization between unpaired electron and the magnitude of ZFS. (C) Partial charge density contours of the two unpaired electrons ("hole" – blue, "electron" – red) for the SW triplet DTEs. Since the unpaired electrons of SW$_{ax}$ are resonant with the CB and VB, they are substantially more delocalized compared to those of SW$_{he}$.

As a result, triplet excitons trapped at the SW$_{he}$ and SW$_{ax}$ defects exhibit substantially different electron density distributions. We modeled the triplet states by constraining the total magnetization of the supercell. As shown in Figure 2A, one of the unpaired electrons of $SW_{he}$ is expectedly stabilized at the formerly unoccupied in-gap defect level and the other one resides at the defect-centered VB maximum state. In the case of $SW_{ax}$, one unpaired electron occupies a state resonant with the valence band, resulting in substantially delocalization.

Electron density localization directly affect the magnitudes of their ZFS, as shown schematically in Figure 2B. In light-element systems such as SWCNTs, the ZFS is primarily caused by the anisotropic magnetic dipole-dipole interaction between the unpaired electrons in the triplet state. This interaction decreases with the cube of the distance between the electrons. Therefore, when the electron density is spread out, the average distance between the electrons increases, reducing the ZFS. Conversely, in triplet systems where the electrons are confined to a small region, their distance is smaller on average, leading to stronger dipole-dipole interactions and larger ZFS parameters. To address this quantitatively, we compute the full spin-dipolar ZFS tensors (D) from the Kohn-Sham electron densities, as outlined in the Methods section. The computed ZFS is then expressed using the conventional parameters $D$ and $E$: $D = D_{zz} - \frac{1}{2}(D_{zz} + D_{yy})$, $E = \frac{1}{2}(D_{yy} - D_{xx})$, where $D_{xx}$, $D_{yy}$, and $D_{zz}$ are the principal values of the ZFS tensor.

First, we compute the ZFS for the triplet state of $SW_{he}$. As shown in Figure 2C, one of its unpaired electrons is stabilized at the formerly unoccupied mid-gap level and the other resides at the defect-centered VB maximum state (cf. Figure 2A). Consequently, both of their electron densities are well-localized around the defect site, leading to a relatively large $D$ value of –473.06 MHz (–1.95 µeV; see Table 1), in good agreement with the experimentally observed triplet in (6,5) SWCNT ($D$ = |345| MHz in Ref. [25]; $D$ = |550| MHz in Ref. [28]). While the sign of $D$ cannot be determined from the reported ODMR measurements, our calculations for $SW_{he}$, which show $D < 0$, indicate that the spin density is stretched along the nanotube axis [55]. The calculated angle between the axis and the principal direction of $D$ is only about 3°. At the same time, our calculations reveal significant deviation of the spin density distribution from axial symmetry, leading to anisotropy of the ZFS tensor ($D_{yy} \neq D_{xx}$) and reflected in a non-zero value of the parameter $E$. Our calculated $E/D$ ratio of 0.26 suggests strong azimuthal anisotropy (rhombicity), which agrees with $E/D$ = 0.23 derived from the ODMR measurements of Ref. [26].

In contrast to the $SW_{he}$ case, $SW_{ax}$ is an example of a system where one of the electrons constituting the triplet resides on the state resonant with the valence band. This state largely retains the character of the VB maximum of a pristine nanotube, resulting in substantial delocalization (cf. Figure 2A,B). The situation where one electron is relatively confined, and the other is spread over the SWCNT direction manifests in a considerably smaller $D$ value around –100 MHz directed along the nanotube axis.

Similarly, we compute the $D$ and $E$ values for the triplet excited states of all other TDs considered in this work (namely, DVs and ADs). The results are listed in Table 1. A noteworthy entry is $DV_{re}$, with its high spin density localization and large ZFS comparable to that of $SW_{he}$. However, since this defect features a deep mid-gap state and relatively high formation energy, we still consider $SW_{he}$ as the most likely candidate for the experimentally observed triplet exciton.

The remaining defects resemble SW$_{ax}$, where one electron is confined at the defect site, while the second remains relatively delocalized. As a result, they exhibit ZFS values in the range of 100–200 MHz. Alongside SW$_{ax}$, these TDs can therefore be considered candidates for the second triplet species (the one with a smaller ZFS of $D = |150|$ MHz) identified in Ref. [28]. By contrast, excitons in pristine (6,5) SWCNTs are expected to be substantially more delocalized than the defect-trapped cases considered in this work (with an estimated exciton size of 13 nm according to Ref. [56]) and are likely to exhibit significantly lower ZFS.

To provide further context, we also compare these ZFS values with those of triplet excitons bound to another widely studied class of SWCNT defects: $sp^3$ defects introduced by covalent doping with aromatic functional groups [15–19]. As a benchmark, we consider a model of a 4-bromobenzene group covalently attached to a (6,5) SWCNT, with a hydrogen atom compensating for the extra electron at an adjacent carbon site, adapted from Ref. [18]. The resulting triplet state exhibits a ZFS $D$ value of approximately –100 MHz, comparable to SW$_{ax}$. This comparison indicates that some TDs, such as SW$_{he}$, facilitate more localized excitons than $sp^3$ defects, making them particularly promising for applications in quantum communication, quantum sensing, and photonic devices. While $sp^3$ defects benefit from well-established implantation techniques via covalent chemistry, TDs represent a compelling alternative, especially for applications where enhanced spin localization is critical.

**Table 1**. DFT-calculated spin-spin ZFS ($D$ and $E$ values along with the angle $\theta$ between the $D_{zz}$ principal direction and the nanotube axis) of triplet excitons trapped by TDs.

| Defect | $D$, MHz* | $E$, MHz | $\theta$, deg |
|---|---|---|---|
| SW$_{he}$ | -473.06 | -123.58 | 3.13 |
| SW$_{ax}$ | -100.81 | -25.49 | 7.86 |
| DV$_{he}$ | 89.90 | 3.68 | 16.18 |
| DV$_{ax}$ | -197.97 | -6.59 | 25.42 |
| DV$_{re}$ | 307.72 | 58.44 | 95.33 |
| AD$_{he}$ | -136.07 | -1.26 | 7.56 |
| AD$_{ax}$ | -107.89 | 23.89 | 91.85 |

*For comparison, the $D$ values obtained from the experimental ODMR spectra in Ref. [28] are $|550|$ MHz and $|150|$ MHz for the excitons labeled T1 and T2, respectively. In Ref. [26], the experimental ODMR spectra were simulated with $D = |345|$ MHz and $E = |78|$ MHz.

After identifying SW$_{he}$ as the most likely exciton trapping site, we aim to study its lowest singlet excited state (labeled S$_1$ in Figure 3A). The elecronic structure of SW$_{he}$ (cf. Figure 2C) suggests that S$_1$ is likely associated with the transition between the defect-centered VB maximum and the in-gap defect state. This was verified using TDDFT calculations performed on a cluster model with atomic positions extracted from the optimized SW$_{he}$–containing supercell (cf. Methods section). An analysis shows that the S$_0$ → S$_1$ excitation is 97% comprised of a single electronic transition between the perturbed VB maximum (highest occupied state) and the in-gap defect state (lowest unoccupied state), indicating that S$_1$ is adequately described by a single Slater determinant. Therefore, we applied the ΔSCF approach [40, 41] to model the excited singlet S$_1$ in the supercell, imposing constrained occupations of the Kohn-Sham states with an electron transeferred from the perturbed VB maximum to the defect state.

Subsequently, we performed atomic relaxation for the excited-state electron configuration with ΔSCF to compute the corresponding zero-phonon line (ZPL). For consistency, we used the same procedure applied throughout the paper – geometry optimization with the PBE functional followed by a single-shot SCF calculation with the hybrid HSE06 functional. The ZPL of the $S_0 \leftrightarrow S_1$ transition calculated this way is 0.77 eV (1610 nm). As an additional check, we performed geometry optimization of both the $S_0$ and $S_1$ states using the HSE06 functional, which yielded nearly the same value of 0.76 eV (1631 nm). These results suggest that the $SW_{he}$ TD is likely to emit in the telecom range (L-band), similar to $sp^3$ defects for which single-photon emission at telecom wavelengths has been experimentally observed [17].

Next, we explored the coupling of the $S_0 \leftrightarrow S_1$ transition to lattice vibrations. The configuration coordinate diagram based on ΔSCF calculations is show in Figure 3A. The effective one-dimensional configuration coordinate $Q$ is defined according to the approach introduced by Alkauskas et al. [54, 57]. The structural change, $\Delta Q$, between the ground and excited states of $SW_{he}$ is moderate, constituting 0.37 amu$^{1/2}$Å (amu = atomic mass units). Subsequently, we determine the Huang-Rhys factor within the one-dimensional approximation as $S_{\mathrm{HR}} = \frac{1}{2\hbar}\Omega\Delta Q^2$, where $\Omega$ is the effective frequency [57, 58]. The computed Huang-Rhys factor is only 0.74, signifying weak electron-phonon coupling and a low number of phonons involved in the transition. Defects with a low Huang-Rhys factor ($S_{\mathrm{HR}} \approx 1$) exhibit a high fraction of photons emitted into the ZPL. These photons are in a well-defined quantum state and are thus useful for quantum information applications [59]. Consequently, our calculations suggest that the $SW_{he}$ defect in (6,5) SWCNT demonstrates the features of a promising telecom-band single-photon emitter.

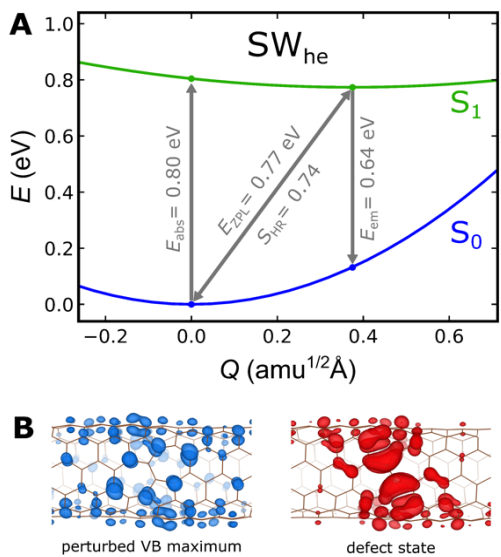

**Figure 3.** (A) Configuration coordinate diagram of the $S_0 \rightarrow S_1$ internal transition of the $SW_{he}$ defect in (6,5) SWCNT obtained from DFT calculations. The grey lines indicate optical transitions, labeled with the corresponding transition energies ($E_{abs}$, $E_{em}$, and $E_{ZPL}$ stand for absorption, emission, and zero-phonon line, respectively). (B) Charge density isosurface of the two states involved in the $S_0 \rightarrow S_1$ transition: the perturbed VB maximum that is occupied in the ground state $S_0$ (*left*) and the defect-induced trap state that becomes occupied in the excited state $S_1$ (*right*).

Another remarkable result of our ΔSCF calculations for SW$_{he}$ is a vanishing energy gap between S$_1$ and the excited triplet state (commonly termed the singlet–triplet gap, Δ$E_{ST}$ = 0.004 eV in our calculations). Small Δ$E_{ST}$ signifies minimal spatial overlap between the orbitals involved in the formation of the S$_1$ state, which minimizes the exchange integral [60, 61]. As illustrated in Figure 3B, a minimal exchange integral is achieved by the localization of the electron densities on alternating atoms of the SW$_{he}$ system. Organic molecules with similar properties are often designed to act as chromophores for organic light-emitting diodes [62], where a reduced singlet–triplet gap is desirable as it facilitates efficient thermally activated delayed fluorescence (TADF) via reverse ISC from the triplet state to S$_1$.

## 4. CONCLUSIONS

In summary, our theoretical results provide new insights into the photophysics of topological defects in semiconducting SWCNTs. The TDs considered in this work only introduce local distortions and do not alter the chirality, yet they can significantly impact the electronic structure of the nanotube and, consequently, exciton dynamics. Our calculations show that TDs enable sufficient electron density localization of the triplet excitonic state to match the experimentally observed range of ZFS values. These findings suggest that the ODMR signals observed in nominally pristine nanotubes are very likely associated with intrinsic defects, with TDs – given the ubiquity – as prominent candidates. Moreover, the possibility of intentionally creating TDs via post-synthetic treatments [32] opens new avenues for engineering the photophysical properties of SWCNTs, particularly for applications in quantum communication and sensing.

Among the studied TDs, SW$_{he}$ exhibits particularly intriguing properties. This defect introduces a shallow trap state into the band gap and causes electron density localization at the valence band maximum. As a result, the unpaired electrons involved in its triplet state are spatially confined, leading to a relatively large ZFS and rendering it a plausible defect to be associated with the triplet exciton observed in Refs. [25, 26, 28]. Notably, the lowest singlet excited state of SW$_{he}$ can undergo a radiative internal transition to the ground state with a ZPL at 1.6 μm (in the telecom L-band) and a low Huang-Rhys factor of 0.74, providing compelling possibilities for applications as single-photon emitters.

**DATA AVAILABILITY**

The data supporting the findings of this study are available within the article. Additional data related to this paper may be requested from the corresponding authors.

**AUTHOR CONTRIBUTIONS**

T.B.: conceptualization, data acuisition and analysis, writing. U.G.: analysis, funding acquisition, review and editing. W.G.S.: conceptualization, funding acquisition, review and editing.

**CONFLICT OF INTEREST**

There are no conflicts to declare.

ACKNOWLEDGEMENTS

We are grateful to Sofie Cambré and Etienne Goovaerts for helpful discussions. We acknowledge Deutsche Forschungsgemeinschaft (DFG) from TRR 332 142/3-2024, Project No. 231447078 for funding as well as the Paderborn Center for Parallel Computing (PC2) for the provided computational resources.